\documentclass[aps,prl,reprint,amsmath,amssymb,superscriptaddress,longbibliography]{revtex4-2}
\usepackage{graphicx}
\usepackage[colorlinks=true,urlcolor=blue,citecolor=blue,linkcolor=blue,bookmarks=false, pdfstartview={FitH}]{hyperref}
\usepackage{amsmath}
\usepackage{bm}
\usepackage{braket}
\usepackage{dcolumn}
\usepackage{textcomp}
\usepackage{physics}
\usepackage{color,soul}
\usepackage{ragged2e}
\usepackage{color,soul}

\begin{document}
\title{Spin-Mediated Direct Photon Scattering by Plasmons in BiTeI
}

\author{A.\,C.~Lee}
\email{aclee@physics.rutgers.edu}
\affiliation{Department of Physics \& Astronomy, Rutgers University, Piscataway, NJ 08854, USA}

\author{S.~Sarkar}
\affiliation{Department of Physics, Concordia University, 7141 Sherbrooke St. W, Montreal, Quebec}

\author{K.~Du}
\affiliation{Department of Physics \& Astronomy, Rutgers University, Piscataway, NJ 08854, USA}
\affiliation{Rutgers Center for Emergent Materials, Rutgers University, Piscataway, NJ 08854, USA}

\author{H.-H.~Kung}
\altaffiliation[Current affiliation: ]{Quantum Matter Institute, University of British Columbia, Vancouver, BC V6T 1Z4, Canada}
\affiliation{Department of Physics \& Astronomy, Rutgers University, Piscataway, NJ 08854, USA}

\author{C.\,J. Won}
\affiliation{Laboratory for Pohang Emergent Materials and Max Planck POSTECH Center for Complex Phase Materials, Department of Physics, Pohang University of Science and Technology, Pohang 37673, Korea}

\author{K.~Wang}
\affiliation{Department of Physics \& Astronomy, Rutgers University, Piscataway, NJ 08854, USA}

\author{S.-W.~Cheong}
\affiliation{Department of Physics \& Astronomy, Rutgers University, Piscataway, NJ 08854, USA}
\affiliation{Rutgers Center for Emergent Materials, Rutgers University, Piscataway, NJ 08854, USA}
\affiliation{Laboratory for Pohang Emergent Materials and Max Planck POSTECH Center for Complex Phase Materials, Department of Physics, Pohang University of Science and Technology, Pohang 37673, Korea}

\author{S.~Maiti}
\email{saurabh.maiti@concordia.ca}
\affiliation{Department of Physics, Concordia University, 7141 Sherbrooke St. W, Montreal, Quebec}

\author{G.~Blumberg}
\email{girsh@physics.rutgers.edu}
\affiliation{Department of Physics \& Astronomy, Rutgers University, Piscataway, NJ 08854, USA}
\affiliation{National Institute of Chemical Physics and Biophysics, 12618 Tallinn, Estonia}

\date{\today}

\begin{abstract}
% Abstract - Proofed %
We use polarization resolved Raman spectroscopy to demonstrate that for a 3D giant Rashba system the bulk plasmon collective mode can directly couple to the Raman response even in the long wavelength $\mathbf q \rightarrow 0$ limit. 
Although conventional theory predicts the plasmon spectral weight to be suppressed as the square of its quasi-momentum and thus negligibly weak in the Raman spectra, we observe a sharp in-gap plasmon mode in the Raman spectrum of BiTeI below the Rashba continuum. 
This coupling, in a polar system with spin-orbit coupling, occurs without assistance from phonons when the incoming photon excitation is resonant with Rashba-split intermediate states. 
We discuss the distinctive features of BiTeI's giant Rashba system band structure that enable the direct observation of plasmon in Raman scattering.
\end{abstract}

\maketitle
% MAIN TEXT 
% State of the Art of the Problem (context, challenge, action, results) - Proofed % 
\paragraph*{Introduction:} 
Plasmons, the collective oscillations of the electron density relative to the nuclei lattice, are longitudinal waves whose energy, $\Omega_{pl}(\mathbf q)$, is finite in 3D systems even in the long wavelength limit where  the quasi-momentum $\mathbf q \rightarrow 0$~\cite{pines2018elementary,mahan1981many}. 
Although Raman spectroscopy can probe optical modes such as phonons, the spectral weight of scattering from bulk plasmons is expected to be vanishingly weak because within the conventional paradigm the Raman intensity is proportional to the square of the quasi-momentum transfer $\mathbf q$~\cite{pines1966,hayes2012scattering}. 

This suppression of the spectral weight for Raman scattering from conventional metals can be understood from the charge conservation perspective. 
In the long-wavelength limit, there are no fluctuations in the total charge, without which no dipole or quadrupole fields that could scatter photons are produced. 
At finite $\mathbf q$, a longitudinal dipole field is produced and Raman scattering being a two photon process couples to it as its square leading to quadrupolar coupling. 
The selection rule that applies to the plasmon-photon coupling dictates that the response would be picked up in the irreducible representations that contain the $x$, $y$, or $z$ fields with strengths $q_x^2$, $q_y^2$ and $q_z^2$. 
Under conventional circumstances, even if the scattering signals were boosted by a resonant Raman process, this $q^2$-suppression still persists~\cite{sarma1999resonantraman}. 
This makes it challenging, if not impossible, to directly observe Raman scattering by a plasmon. 

The earlier studies~\cite{pinczuk1971resonant,Klein1983,pinczuk1989large,goni1991onedimensional,sarma1999resonantraman} that were able to couple to a plasmon by Raman process involved indirect help from extrinsic sources, e.g., a technique that couples the plasmon to a boson of similar energy at finite momentum to produce a polariton was applied to observe a plasmon in Raman spectra, particularly for low-carrier concentration semiconductors~\cite{varga1965coupling,mooradian1966observation,mooradian1967polarization,abstreiter1979coupled}. 
The application of a high magnetic field has also been shown to produce observable Raman collective charge-density fluctuations~\cite{Slusher1967,Slusher1968} such as in the two-dimensional fractional quantum Hall systems~\cite{girvin1985collective,girvin1986magneto,pinczuk1993observation,pinczuk1994inelastic,eriksson1999collective}. 

In systems with inversion, there is no fully symmetric intrinsic vector that could produce a dipole field and hence a finite $q$ is required for Raman coupling to light. 
In prior studies, the visibility of the plasmon was enhanced either by periodic structures that break the translational symmetry and inversion or by borrowing the spectral weight from other bosons with which the plasmon hybridises~\cite{varga1965coupling,mooradian1966observation,mooradian1967polarization,platzman1965incoherent,SlusherPlasmons1968}. 
The visibility can also be enhanced in highly absorptive materials where $q$ is not a good quantum number for plasmon excitation~\cite{Cardona1984}, or by being close to the threshold of the particle-hole continuum from where some spectral weight could be borrowed~\cite{SlusherPlasmons1968,platzman1965incoherent}. 
In all other known cases, the $q^2$-suppression renders the plasmon invisible. 
There is one more instance where bare plasmons could be made visible, which requires the LO-TO splitting of phonons in high-symmetry polar systems, but without actually hybridizing with the phonons. 
In this mechanism, it is the polar oscillation of the phonons that provides the dipolar field necessary to excite the plasmon, however, such excitations would only be visible in a cross-polarization geometry~\cite{mooradian1967polarization,Klein1983}.  

In polar systems with broken inversion, one can also have strong spin-orbit coupling. 
It then opens up the question of if the photon scattering from plasmons could have a different weight from $q^2$ in such systems, without any help from phonons, leading to a potential direct observation of plasmons in Raman scattering from a purely electronic mechanism.

% Basic Properties Figure - Proofed %,
\begin{figure}[t] 
	\includegraphics[width=0.95\linewidth]{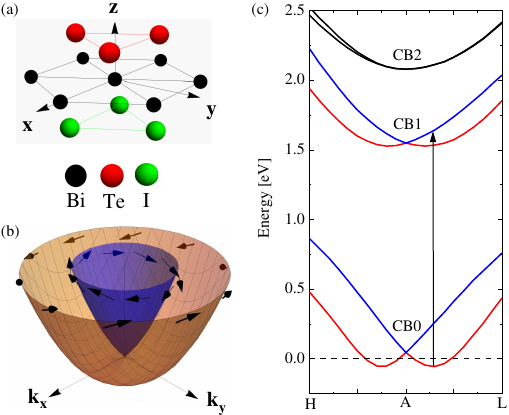} 
		\vspace{-2mm}    
	\caption{ \textbf{Basic properties of BiTeI.} 
		(a) The crystal structure. 
        (b) Spin texture of Rashba spin-split bands. 
        The inner and outer Rashba bands have opposite helicities for the same in-plane quasi-momentum direction. 
        The orbital Rashba effect results in slight z-canting of the spins. 
        (c) An illustration of the energy-band dispersion relative to the Fermi energy based on a DFT calculation in the Ref.~\cite{ishizaka2011giant}. 
         The red (blue) band color-coding indicate distinct spin-polarized bands which are split by the Rashba interaction. 
         The black bands (CB2) are unpolarized. 
        The vertical arrow from CB0 to CB1 band shows electronic transitions responsible for resonance enhancement of the plasmon Raman intensity. 
    \vspace{-5mm}    
      }
 	\label{basic}%
\end{figure}

%\paragraph*{Choice of material:} 
The observation above is the prime motivation to study the 3D giant Rashba polar material, BiTeI~\cite{ishizaka2011giant}. 
The unit cell has Bi$^{3+}$ atoms that are sandwiched between layers of I$^{-}$ and Te$^{2-}$ atoms forming a tri-layer of atoms [space group P3\textit{m}1 (C$_{3v}$)] stacked along the c-axis, Fig.\,\ref{basic}(a), which gives it a non-centrosymmetric nature and is the origin of the orbital Rashba effect which intrinsically couples the charge and spin degrees of freedom without the need of an external gating or surface-induced symmetry breaking. 
The different tri-layers are van der Waals-bonded strong enough to result in a 3D band structure. 
This results in spin-split helical electronic bands~\cite{park2011orbitalangularmomentum} with the spins also showing a slight out-of-plane canting, Fig.\,\ref{basic}(b). 
Although expected to be an insulator, the I$^{-}$ vacancies that occur during the crystal growth cooling process raise the chemical potential into the lower subbands~\cite{tomokiyo1977phase} [labelled as CB0 in Fig.\,\ref{basic}(c)], resulting in a metallic character with low carrier concentrations, $n_c$, such that the out-of-plane plasma frequency is expected to be in 60-80\,meV range~\cite{lee2011optical}. 
The self-doping is such that chemical potential remains below the Dirac-like point. 
This aspect introduces a gap \textit{below} the inter-sub-band excitations~\cite{ye2015transport,lee2022chiral} which will play a central role in the data interpretation. 

\begin{figure}[t]%
	%\centering  \captionsetup{justification=RaggedRight}
	\includegraphics[width=.85\linewidth]{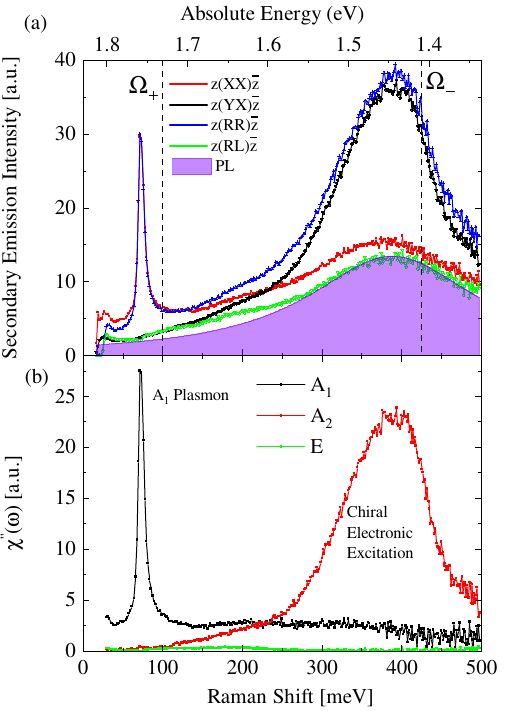}%
		\vspace{-2mm}    
	\caption{\textbf{Secondary emission and Raman response of BiTeI.} 
	(a) Secondary emission spectra taken from sample A in the \textit{z}(RR)$\bar{z}$ (blue), \textit{z}(RL)$\bar{z}$ (green), \textit{z}(XX)$\bar{z}$ (red), and \textit{z}(YX)$\bar{z}$ (black) scattering geometries using low-resolution settings. 
	The vertical dotted lines represent the bounds of the Rashba continuum, $\Omega_{\pm}$~\cite{lee2022chiral}. 
        (b) Symmetry decomposed Raman spectra into $A_{1}$, $A_{2}$, and $E$ channels~\cite{Note1}. 
        No noticeable $E$-symmetry excitations where detected within this frequency range.
        All measurements were taken at 15\,K using $\omega_{i}$ = 1.83\,eV excitation energy. 
        The upper scale denotes absolute energy of emission $\omega_{s}$ and the lower scale denotes Raman shift energy $\omega = \omega_{i} - \omega_{s}$. 
    \vspace{-5mm}    
 	}
	\label{SecEmission}%
\end{figure}

\paragraph*{Main finding:}
In this Letter, we report the unexpected observation of a distinct plasmon mode in the Raman spectrum of BiTeI, a material with a giant Rashba spin-orbit coupling (SOC). 
In contrast to conventional expectations, our findings reveal strong and direct light scattering by bulk plasmons even in the long-wavelength limit. 
This observation challenges the prevailing understanding of plasmon-assisted Raman processes and highlights the unique features of the giant Rashba system in BiTeI.

\begin{figure}[t]%
	\includegraphics[width=0.95\linewidth]{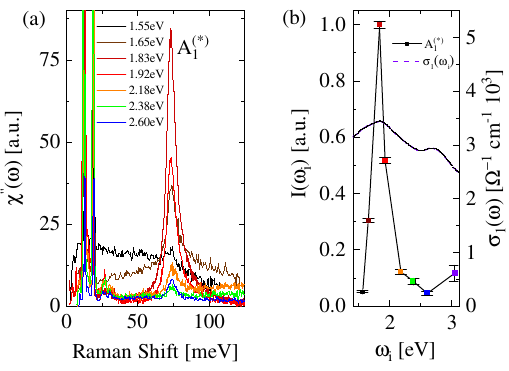}
	\vspace{-2mm}    
	\caption{\textbf{Resonant Raman excitation profile (RREP) of the $A_{1}^{*}$ mode.} 
        (a) The  Raman response in the fully symmetric symmetry channel from the sample A taken in \textit{z}(XX)$\bar{z}$ scattering polarization for $\omega_{i}=$1.55--2.60\,eV excitation energies. 
        The sharp low-frequency modes are fully-symmetric phonons. 
        (b) The RREP - spectral intensity of the $A_{1}^{(*)}$ mode integrated over Raman shift and plotted against $\omega_{i}$. 
	    The real part of the optical conductivity, $\sigma_{1}(\omega_{i})$, is overlaid for reference. 
     \vspace{-5mm}    
        }
	\label{PPExcit}%
\end{figure}

The plasmon collective mode appears at around 74\,meV, which is below the inter-sub-band continuum of excitations, in the fully symmetric channel of the Raman spectrum, see Fig.\,\ref{SecEmission}. 
By studying the resonant Raman excitation profile (RREP), Fig.\,\ref{PPExcit}, 
we observe that the scattering intensity is induced by resonant electronic transitions between mutually spin-split electronic bands CB0 and CB1, see Fig.\,\ref{basic}(c), but not between bands where there is no mutual spin splitting like between CB0 and CB2. 
By studying different samples from the same batch with varying carrier density we show that the square of the mode's energy scales linearly with the carriers concentration $n_{c}$, thereby confirming the plasmonic nature of the mode. 
We also track the spectral lineshape of the plasmon as a function of temperature and show that its relaxation rate is consistent with the behavior of an in-gap collective mode. 

% Excitation Profile Figure - Not Proofed %

\paragraph*{Experimental details:} 
Single crystals of BiTeI were grown using the vertical Bridgman technique~\cite{sankar2014roomtemperature,kanou2013crystal}. 
Four-probe Hall effect measurements were performed using a Physical Property Measurement System (PPMS) by Quantum Design to determine the $n_{c}$ of the samples; see~\footnote{See Supplementary Materials at [URL will be inserted by publisher] for the details of crystal growth and characterization, measurement methods, analysis of incident angle dependent Raman data, Raman selection rules, analysis of Raman data temperature dependence, which includes Refs. \cite{sankar2014roomtemperature,kanou2013crystal,hurd2012hall,goldstein2002classical,loudon1965theory,blumberg1994investigation,makhnev2014optical,dresselhaus2008applications,dresselhaus2008applications,basov2005electrodynamics,ishizaka2011giant}} for full experimental details. 
The samples used in this work are denoted as A, B, C, D, and E. 
The crystals were exfoliated in a nitrogen-rich environment and then transferred into a continuous He-flow optical cryostat. We use polarization-resolved Raman scattering measurements in the quasi-back-scattering geometry. 
Scattering polarization geometries, defined by the incident (collected) light polarizations, e$_{i(s)}$, and light propagation directions relative to the crystal structure, $\hat{k}_{i(s)}$, will be denoted as $\hat{k}_{i}$(e$_{i}$e$_{s}$)$\hat{k}_{s}$, where $\hat{k}_{i}=-\hat{k}_{s}$. 
The linear light polarization directions, X and Y, lie within the \textbf{xy}-plane and are orthogonal to one another; due to the symmetries of $C_{3v}$, the choice of in-plane linear light polarization directions does not affect the results. 
R and L denote right and left circularly polarized light, respectively, such that R(L) = X $\pm i$Y~\cite{Note1}.  
Finally, the energy of the incident (collected) photons will be denoted as $\omega_{i(s)}$, respectively with $\omega = \omega_{i} - \omega_{s}$ being the Raman shift. 
The Raman susceptibility will be denoted as $\chi^{''}(\omega)$. 

\paragraph*{Data and interpretation:} % Secondary Emission Figure - Proofed % 
In Fig.\,\ref{SecEmission}(a) we show spectra of secondary emission from the sample A measured in four in-plane  polarization geometries. 
The broad feature centered at $\omega_{s}=1.48$\,eV energy (shaded in violet) is photo-luminescence (PL) which appears in all polarization channels~\cite{lee2022chiral}. 
By subtracting the PL and decomposing the remaining signal into symmetrized Raman response channels, see~\cite{Note1} for details, several excitation features become evident. 
In Fig.\,\ref{SecEmission}(b), we show that a sharp in-gap collective mode appears at 74\,meV in the $A_{1}$ symmetry channel that we denote as $A_{1}^{*}$. 
In addition, there are two electronic continua of excitations:
the $A_{1}$ continuum that extends from just above the $A_{1}^{*}$ mode to 500\,meV and the $A_2$ Rashba continuum, which was studied in Ref.\,\cite{lee2022chiral}, that is an order of magnitude stronger than the $A_{1}$ continuum. 
No appreciable signal was collected in the E symmetry channel. 

The $A_{1}^{*}$ mode is the main finding of this work. 
This mode cannot be an optical phonon as all the phonons lie at lower frequencies and have been accounted for~\cite{ponosov2014dynamics,gnezdilov2014enhanced,tran2014infrared,sklyadneva2012lattice}. 
Likewise, the mode cannot be a plasmon-polariton as the energy difference between the plasmon energy and the longitudinal optical modes is too great for reasonable coupling~\cite{varga1965coupling}. 
Although bulk plasmon scattering in Raman spectroscopy is typically treated as forbidden, we attribute the mode to photon-bulk plasmon scattering for reasons discussed below. %that are described in~\cite{surajit2023plasmon}. 

% Reasoning for Excitation Profile %
We acquired Raman spectra of the $A_{1}^{*}$ mode as a function of excitation energy $\omega_{i}$ to track its spectral weight. 
In Fig.\,\ref{PPExcit}(a), we show the Raman response of BiTeI using a set of laser excitation energies between 1.55\,eV and 2.60\,eV. 
In Fig.\,\ref{PPExcit}(b) we plot the RREP of the $A_{1}^{*}$ mode, $I(\omega_{i})$, where the real part of the optical conductivity, $\sigma_{1}(\omega_{i})$, is also overlaid for reference. 
The RREP of the $A_{1}^{*}$ mode peaks at $\omega_{i}=1.83$\,eV and is nearly undetectable away from this sharp resonance excitation condition. 
Based on the RREP and calculations of the electronic band structure of BiTeI, the resonance enhancement of the mode is a result of transitions from the occupied CB0 band to intermediate CB1 band state, see Fig. \ref{basic}(c). 

% Temperature Dependence Figure - Not Proofed %
\begin{figure}[t]
	%\centering\captionsetup{justification=RaggedRight}
	\includegraphics[width=0.85\linewidth]{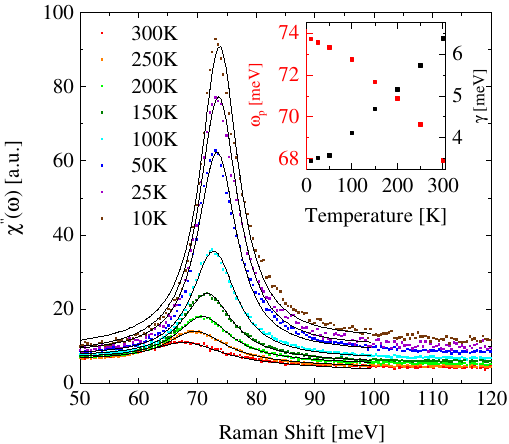}
		\vspace{-2mm}    
	\caption{\textbf{Raman response of BiTeI as a function of temperature.} 
	The Raman response for the sample A taken in \textit{z}(XX)$\bar{z}$ scattering geometry at temperatures between 10 to 300\,K using $\omega_{i}=1.83$\,eV resonant excitation energy. 
	(inset) The energy and HWHM values derived from Lorentzian fits to the $A_{1}^{*}$ mode lineshape accounting for spectral resolution.
    \vspace{-5mm}    
 	} 
	\label{PPTemperature}
\end{figure}

% Temperature Investigation %
We also investigated the temperature dependence of the $A_{1}^{*}$ mode near its excitation resonance. 
In Fig.\,\ref{PPTemperature}, we show the Raman response for 10--300\,K temperatures range. 
In the inset of Fig.\,\ref{PPTemperature}, we plot the energy, $\omega_{p}(T)$, and the half-width-at-half-maxima (HWHM), $\gamma(T)$, of the mode as a function of temperature. 
The spectral lineshape monotonically red-shifts and sharpens upon cooling. 
These results are consistent with the picture of a $T$-dependent damped oscillator. 
However, the lineshape of the $A_{1}^{*}$ mode is slightly asymmetrical, suggesting that an additional relaxation contributes to damping of the in-gap collective mode.

\begin{figure}[t]%
	\includegraphics[width=0.98\linewidth]{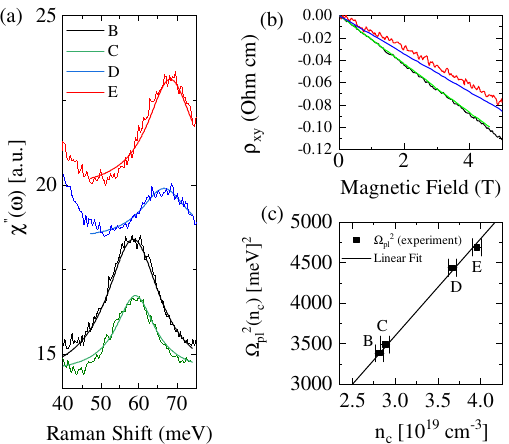}
	\caption{\textbf{The relationship between the square of the $A_{1}^{*}$ mode energy and $n_c$.} 
        (a) Secondary emission spectra of the $A_{1}^{*}$ mode for samples B, C, D, and E taken at room temperature. 
        The collective mode is fitted to a Lorentzian-like function~\cite{surajit2023plasmon} after background subtraction and shifted vertically by a constant proportional to $n_c$ of the sample to aid the reader's eye.
        (b) Transverse resistivity data for the samples listed in (a). 
        The data has been normalized to account for sample dimensions. 
        (c) $\Omega_{pl}^{2}$ plotted against $n_{c}$. 
        The solid black line is the line-of-best-fit for the data points fixing the y-intercept to be zero. 
    %\vspace{-5mm}    
 	}
	\label{halleffect}
\end{figure}

\paragraph{Analysis and justification:}
The selective presence of the mode in the $A_1$ channel establishes the charge nature of the excitation and thus rules out possible chiral spin modes expected in Rashba metals, as they would appear in the $A_2$ symmetry channel~\cite{kung2017chiral}. 
We repeated the same experimental procedure for samples B, C, D, and E which had different concentration of I$^{-}$ vacancies. 
We found the carrier densities for all the samples using Hall measurements, Fig.\,\ref{halleffect}(b), and tracked the energy of the $A_1^*$ mode, Fig.\,\ref{halleffect}(a). 
Upon plotting the square of the mode's energy as function of the carrier concentration $n_{c}$, Fig.\,\ref{halleffect}(c), we observe a straight line confirming its plasmonic nature~\cite{cardona2005fundamentals}. 
The solid black line is the best fit to the data points with the constraint $\Omega_{pl}^{2}(0) = 0$. 
Within parabolic electron dispersion model with out-of-plane effective mass $m_c^*$, the slope of this line is given by $4 \pi e^2 \hbar^2/m_c^*\epsilon_\infty=1.2\times10^{-16}$\,[meV$^2$-cm$^{3}$]. 
The derived value of $m_c^* = 1.02m_e$, where $m_e$ is the mass of a free electron, well agrees with the value acquired by the optical reflectivity measurements~\cite{lee2011optical}. 
We have assumed that for the c-axis direction $\epsilon_\infty=12.2$, which was found from grazing-angle reflectivity measurements~\cite{lee2011optical}. 

% Hall Effect Measurements Figure - Proofed %

The sharp resonant enhancement of the plasmon scattering cross-section which is shown in the RREP plot of the Fig.\,\ref{PPExcit}(b) requires further explanation. 
The energy gap between CB0 and CB1 bands, see Fig. \ref{basic}(c), can be discerned from the broad maxima in the $\sigma_{1}(\omega_i)$ spectra. 
Although there is an additional broad maximum in $\sigma_{1}(\omega_{i})$ at $\omega_{i} \sim 2.7$\,eV, the $A_{1}^{*}$ mode's RREP does not have second enhancement resonance peak at this excitation energy. 
This suggests that the CB2 bands, which are not Rashba spin-split, play no part in the resonance enhancement of the $A_{1}^{*}$ mode. 
Since both the CB0 and CB1 bands are Rashba spin-split~\cite{ishizaka2011giant}, the physical mechanism that allows for the observation of the plasmon mode must require the Rashba spin splitting of the intermediate states as well.

There is another important aspect of the observed plasmon mode that requires a discussion. 
One usually expects for the plasmon to appear at frequencies \textit{above} the continuum of particle-hole excitations. 
From Fig.\,\ref{SecEmission}(b) we observe that the plasmon is an in-gap mode below the continuum. 
This apparent discrepancy is resolved once we note that the continuum here is the chiral continuum of spin-flip excitations~\cite{lee2022chiral}. 
The out-of-plane plasmon is still above the charge sector particle hole continuum that is not visible in Raman spectrum due to $\mathbf q\rightarrow0$ limit. 

It is known that in 2D Rashba metals at finite $\mathbf q$ the plasmon can enter and be damped by the spin-flip continuum~\cite{pletyukhov2006screening}, while for 3D Rashba metals the out-of-plane plasmon does not couple to this spin-flip continuum~\cite{maiti2015collective}. 
The novel feature of this plasmon is that it couples to and is damped by the proximity to the chiral spin-flip continuum~\cite{surajit2023plasmon}. 
This is evident from the weak scattering signal in the continuum region of the $A_1$ channel, and also from the temperature evolution of the scattering rate of the $A_{1}^{*}$ mode. 
What distinguishes BiTeI from 2D Rashba metals [which only models the in-plane spin-orbit coupling (SOC)] is the presence $z$-canting of spins which is about 10\% of the in-plane spin~\cite{bawden2015hierarchical,maass2016spin}. 
We attribute the sensitivity of the $A_1$ response to the spin-flip continuum, and hence the coupling between plasmon and the spin-flip continuum, to the presence of the $z$-canting in the 3D Rashba system band structure. 
The details of this coupling are presented in a concomitant work that explores the observation of plasmons in SOC systems on more general grounds~\cite{surajit2023plasmon}. 
Because the $z$-canting is small, we expect the plasmon to remain well resolved even if it entered the spin-flip continuum. 

% Group Theory %
From a group theoretical perspective, the in-plane plasmons couple to the $E$ representation and the out-of-plane plasmon to the $A_{1}$ representation of the C$_{3v}$ point group, but both still couple with spectral weight $q^2$ without SOC. 
Since BiTeI is a polar system where the out-of-plane dipole ($\bf{z}$) and the in-plane quadrupole ($x^{2}+y^{2}$) excitation transform as the same fully-symmetric $A_1$ representation, they mix allowing the plasmon to appear in the $A_1$ scattering channel. 
The finite spectral weight (non-$\mathbf q^2$) for the coupling is the main discovery of this study and is attributed to the $z$-canting of the spin states~\cite{surajit2023plasmon}.  

% Conclusion - Done %
\paragraph*{Conclusions:}
In this work, we have presented polarization-resolved Raman spectroscopy measurements revealing an unexpected and distinctive plasmon mode in the Raman spectrum of the 3D giant Rashba metal BiTeI. 
The observation challenges conventional expectations of vanishingly weak plasmon intensity in Raman scattering, especially in the long-wavelength limit.
%we have demonstrated that the bulk plasmon appears in the Raman spectra of the 3D giant Rashba metal, BiTeI. 

The tell-tale signatures of $\sqrt{n_c}$-dependence, $A_{1}$ symmetry, and temperature dependence of the plasma frequency and linewidth are present. 
These results are supported by the fact that the plasmon energy is below the Rashba continuum of the single-particle excitations, thus the mode is not overdamped. 
The excitation profile of the plasmon demonstrates that observation of the collective mode is possible only when the laser excitation energy is resonantly tuned to the transition between spin-split by Rashba interaction bands. 
This discovery sheds light on the important role of inversion breaking in the manifestation of special effects in optical spectroscopy and provides a non-phonon, spin-orbit coupling assisted mechanism for photon-plasmon coupling.
Two important questions arise from these results: (1) why do the initial and intermediate states need to be Rashba spin-split and (2) why is resonance needed to observe the plasmon? 
In a concomitant study~\cite{surajit2023plasmon}, we delineate how the coupling between spin-split Rashba bands in the resonant Raman effect could dominate over the conventional $q^2$-coupling in Rashba systems.

% Acknowledgements %
We thank Dmitrii Maslov, Bo Peng, Zuzhang Lin and Bartmeu Monserrat for discussions. 
The spectroscopic work at Rutgers (A.C.L., H.-H.K. and G.B.) was supported by the NSF under Grant No. DMR-2105001. 
The characterization work (K.D. and S.-W.C) was supported by the DOE under Grant No. DOE: DE-FG02-07ER46382. 
The crystal growth at Postech was supported by the National Research Foundation of Korea (NRF) funded by the Ministry of Science and ICT, grant no. 2020M3H4A2084417. 
S.S. and S.M. acknowledge support from the Natural Sciences and Engineering Research Council of Canada (NSERC) Grant No. RGPIN-2019-05486. 
The work at NICPB (G.B.) was supported by the European Research Council (ERC) under the European Union’s Horizon 2020 research and innovation programme Grant Agreement No. 885413.

%
%\bibliography{new_master_bibliography.bib}

\end{document}